\documentclass{pasj01}
\Received{$\langle$reception date$\rangle$}
\Accepted{$\langle$acception date$\rangle$}
\Published{$\langle$publication date$\rangle$}

\usepackage{color}
\usepackage{bm}

\begin{document}

\title{Impulsive Gas Fueling to Galactic Center in a Barred Galaxy Due to Falls of Gas Clouds}

\author{Hidenori Matsui${}^{1,2}$}
\author{Toshiyasu Masakawa${}^{3}$}
\author{Asao Habe${}^{4}$}
\author{Takayuki R. Saitoh${}^{5}$}
\altaffiltext{1}{National Institute of Technology, Asahikawa College, Shunkodai 2-2-1-6, Asahikawa, Hokkaido, 071-8142, Japan}
\altaffiltext{2}{International Centre for Radio Astronomy Research (ICRAR), M468, The University of Western Australia, 35 Stirling Hwy, Crawley, Western Australia, 6009, Australia}
\altaffiltext{3}{The Open University of Japan}
\altaffiltext{4}{Graduate School of Science, Hokkaido University, Kita 10 Nishi 8, Kita-ku, Sapporo, Hokkaido 060-0810, Japan}
\altaffiltext{5}{Department of Planetology, Graduate School of Science, Kobe University, 1-1 Rokkodai-cho, Nada-ku, Kobe, Hyogo 657-8501, Japan}
\email{matsui@asahikawa-nct.ac.jp}


\maketitle

\begin{abstract}
We have studied the evolution of the central hundred pc region of barred galaxies by performing numerical simulations realizing multi-phase nature of gas.
Our simulations have shown that a stellar bar produces an oval gas ring namely the $x$-2 ring within $1~{\rm kpc}$ as the bar grows.
The ring is self-gravitationally unstable enough to trigger formations of gas clouds.
Although the gas clouds initially rotate in the $x$-2 ring, cloud-cloud collisions and/or energy injections into the gas ring by Type II supernovae deviate some of the clouds from the ring orbit.
After the deviation, the deviated clouds repeat collisions with the other clouds, which rotate in the $x$-2 ring, during several rotations.
These processes effectively reduce the angular momentum of the deviated gas cloud.
As a result, the gas cloud finally falls into the galactic center, and episodic gas supply to the galactic center takes place.
\end{abstract}


\section{Introduction}
Barred galaxies generally possess a large amount of gas within their central kpc regions.
In the Milky Way, $<10~\%$ of total molecular gas exists in the central region which is called the Central Molecular Zone (CMZ) (e.g., \citet{1996ARA&A..34..645M,1998A&A...331..959D}).
Gas fueling from the galactic central kpc scale to the pc scale would play an important role in growing the masses of the galactic central supermassive black holes (SMBHs), powering active galactic nuclei (AGNs), or nuclear starburst.
Then, how to transfer gas from the galactic central kpc region to the pc region is a key problem for understanding of the galactic nuclear evolution.

A stellar bar potential is expected to have possibility to remove the angular momentum of gas and transfer the gas to the central pc region.
Previous numerical simulations of barred galaxies, however, have revealed that the inner Lindblad resonance (ILR) barrier prevents the gas from losing its angular momentum and accreting to the galactic center \citep{1994ApJ...425L..73E,2015ApJ...806..150L}.
In order to overcome the barrier, some ideas have been proposed by previous numerical simulations.
For examples, self-gravity instability of gas \citep{1992MNRAS.258...82W,1995MNRAS.277..433W}, double bar structures \citep{1989Natur.338...45S,1993A&A...277...27F,2009ApJ...691.1525N}, or a nuclear LR (NLR) \citep{1998MNRAS.295..463F,2000ApJ...529..109F} have been proposed.
Although their simulations succeed to overcome the barrier, they could not resolve complicated natures of interstellar medium (ISM) in the central region of barred galaxies due to their limited resolutions.

Previous observations of barred galaxies have indicated complicated natures of ISM in their central regions.
In CMZ, observations of giant molecular clouds (GMCs) have detected the signature of clouds-clouds collisions \citep{2015PASJ...67..109T,2018ApJ...859...86T,2020ApJ...903..111T,2021PASJ...73S..75E}.
Furthermore, star formations are suppressed in CMZ in comparison with the other environment such as solar neighborhood although dense molecular gas is abundant \citep{2017A&A...603A..90K}.
In the galactic central region of M83, observations of individual molecular clouds have shown that some clouds are along the $x$-1 and $x$-2 orbits and inside the $x$-2 orbit \citep{2019ApJ...884..100H,2021MNRAS.505.4310C}.
These observations imply star formation induced by could-cloud collisions \citep{2019ApJ...884..100H}.
Dense molecular gas associated with a starburst ring has been observed at the $x$-2 orbit in NGC 6951 \citep{1999ApJ...511..157K,2013A&A...551A..81V}.
In such $x$-2 ring, it has been indicated that fragmentations of gas to GMCs occur due to gravitationally instability of the gas ring in NGC 1097 \citep{2011ApJ...736..129H}.
Thus, in order to investigate evolution in the galactic central region, simulations which are able to resolve cold and dense molecular gas and GMCs are inevitable.

Recently high resolution simulations of a barred galaxy with sub-pc resolution have been performed (e.g., \citet{2015MNRAS.446.2468E,2019MNRAS.488.4663S,2020MNRAS.499.4455T}).
These simulations focus on the Milky Way and succeed to reproduce the size and gas dynamics of CMZ of which size is approximately $200~{\rm pc}$.
Since a central gas ring with larger size, which is $0.5-1~{\rm kpc}$, has been observed in the other galaxies, for examples M83 \citep{2021MNRAS.505.4310C}, NGC 6951 \citep{2013A&A...551A..81V}, and NGC1097 \citep{2011ApJ...736..129H}, our simulations adopt the different external potentials, namely the different galaxy model, from the previous simulations in order to investigate gas dynamics in general barred galaxies rather focusing on the Milky Way. 
Our galaxy models produce the larger size of a central ring, which enables us to elucidate the evolution of a gas ring in the other barred galaxies.
\citet{2021ApJ...914....9M} and \citet{2022ApJ...925...99M} have performed simulations of a larger central ring to study its evolution and star formation activity by assuming various inflow rates to the ring.
On the other hand, our simulations cover the whole of a galaxy instead of assuming various gas inflow rates.
Our numerical simulations of the whole galaxy get time dependent gas inflow rate to the central region, which allows us to understand time evolution of the gaseous ring more realistically.

In this paper, we have performed numerical simulations, which realizes multi-phase nature of interstellar medium by high resolution, of gas in a barred galaxy.
We describe our simulation model method in \S \ref{methods}, results of our simulations in \S \ref{results}, discussion about comparison of our results with observation, spatial resolution, and gas fueling mechanism during bar growth in \S \ref{discussion}, and conclusion in \S \ref{conclusion}.

\section{Method}~\label{methods}

\subsection{Galaxy models}~\label{models}

In our simulations, stellar and dark matter components are given by the external gravitational potential.
We assume a Milky Way-like potential used by \citet{2008PASJ...60..667S}.
In this model, the Miyamoto-Nagai model \citep{1975PASJ...27..533M} and the NFW profile \citep{1996ApJ...462..563N} are adopted as the stellar disk and the dark matter halo, respectively.
These density profiles are as bellows:

\begin{equation}
\rho _{\rm MN} (R,z) = \Big(\frac{b_{\rm d}^2 M_{\rm d}}{4\pi} \Big)
	  \frac{[aR^2 + (a_{\rm d}+3\zeta)(a_{\rm d}+\zeta)^2]}{\zeta^3 [R^2 + (a_{\rm d}+\zeta)^2]^{5/2}}
\end{equation}

\noindent and

\begin{equation}
\rho _{\rm NFW} (r) = \frac{\rho _{\rm c}}{(r/r_{\rm s})(1+r/r_{\rm s})^2},
\end{equation}

\noindent where $M_{\rm d}$, $a_{\rm d}$, $b_{\rm d}$, $\rho _{\rm c}$, and $r_{\rm s}$ are the disk mass, a radial scale length of the disk, a vertical scale length of the disk,
the characteristic density of the halo, and a scale radius of the halo,
respectively.
$R$ and $z$ are a radial distance and a height, respectively, in the Galactocentric cylindrical coordinate system $(R,\phi,z)$.
$r$ is a spherical radius $r=\sqrt{R^2+z^2}$ and $\zeta = \sqrt{z^2 + b_{\rm d}^2}$.
We set to be $M_{\rm d}= 4.0\times 10^{10}~M_{\odot}$, $a_{\rm d}=3.5~{\rm kpc}$, and $b_{\rm d}=400~{\rm pc}$.
By setting to be the virial mass $M_{\rm vir} = 10^{12}~M_{\odot}$ and the concentration parameters $c_{\rm NFW} = 12$,
$\rho _{\rm c}=4.87\times 10^6~M_{\odot}~{\rm kpc^{-3}}$ and $r_{\rm s} = 21.5~{\rm kpc}$ are obtained (see \citet{2008PASJ...60..667S}).

Since this model does not take a stellar bulge and a central supermassive black hole (SMBH) into account,
the bulge and the SMBH are added to the model.
The density profile of the bulge is given by the Plummer model \citep{1911MNRAS..71..460P}:

\begin{equation}
\rho _{\rm b} (R,z) = \frac{3M_{\rm b}}{4\pi a_{\rm b}^3}\Big( 1 + \frac{r^2}{a_{\rm b}^2} \Big)^{-5/2}
\end{equation}

\noindent where $M_{\rm b}$ and $a_{\rm b}$ are the mass and the scale radius of the bulge, respectively.
The potential of the SMBH is given by

\begin{equation}
\Phi _{\rm SMBH} (r) = -\frac{GM_{\rm SMBH}}{\sqrt{r^2+\epsilon _{\rm SMBH} ^2}},
\end{equation}

\noindent where $M_{\rm SMBH}$ and $\epsilon _{\rm SMBH}$ are the mass and the gravitational softening length of the SMBH.
These parameters are set to be $M_{\rm b} = 1.7\times 10^{10}~M_{\odot}$, $a_{\rm b} = 0.5~{\rm kpc}$ \citep{2012PASJ...64...75S}
and $M_{\rm SMBH} = 4.3\times 10^{6}~M_{\odot}$ \citep{2017ApJ...837...30G},
which would create the Milky Way-like potential of bulge and SMBH.
The softening length of SMBH is set to be $\epsilon _{\rm SMBH} = 1 ~{\rm pc}$.

We embed a bar potential into the above axisymmetric potential.
The bar potential is given by the following three dimensional quadrupole model \citep{2016MNRAS.461.3835M}:

\begin{equation}\label{bar}
   \Phi _{\rm bar} (R,\phi,z) = \alpha \frac{v_0^2}{3} \Big( \frac{R_0}{R_{\rm bar}} \Big)^3 U(r) \frac{R^2}{r^2}\cos 2\phi
\end{equation}

\noindent and

\begin{equation}\label{bar2}
   U(r) =
 \left\{ \begin{array}{ll}
   (r/R_{\rm bar})^3 -2 & (r<R_{\rm bar})\\
   -(R_{\rm bar}/r)^3  & (r\ge R_{\rm bar})
 \end{array} \right.
\end{equation}

\noindent where $\alpha$, $R_0$, $v_0$, and $R_{\rm bar}$ are the amplitude of the bar, the radial distance of Sun, the circular velocity at $R_0$, and the length of the bar, respectively.
Here, we use these parameters as $\alpha = 0.01$, $R_0 = 8~{\rm kpc}$, and $R_{\rm bar} = 3.5~{\rm kpc}$,
and the bar pattern speed is set to be $\Omega _{\rm bar} = 52.2~{\rm km/s/kpc}$ similarly to \citet{2000AJ....119..800D} and \citet{2016MNRAS.461.3835M}.
In our galaxy model, $v_0=200~{\rm km/s}$.
We use the rotating coordinate system with the bar pattern speed.
Therefore, $\phi = (\phi ' - \Omega _{\rm bar} t)$, where $\phi '$ is the angle from $x$-axis in the rest frame.
In order to prevent gas from unrealistic behavior, the bar gradually grows from $200~{\rm Myr}$ to $500~{\rm Myr}$ by multiplying the equation (\ref{bar}) by the following factor $\eta (t)$:

\begin{equation}
\eta (t) = \Big( \frac{3}{16}\xi ^5 - \frac{5}{8} \xi ^3 + \frac{15}{16} \xi + \frac{1}{2} \Big),
\end{equation}

\noindent and

\begin{equation}
  \xi = \left\{ \begin{array}{ll}
    -1 & (0\leq t < 200~{\rm Myr}) \\
    \displaystyle \frac{2(t-200~{\rm Myr})}{300~{\rm Myr}} -1 & (200~{\rm Myr} \le t < 500~{\rm Myr}) \\
    1 & (t \ge 500~{\rm Myr})
  \end{array} \right.
\end{equation}

\noindent \citep{2016MNRAS.461.3835M}.

\begin{figure}
 \begin{center}
  \includegraphics[width=80mm]{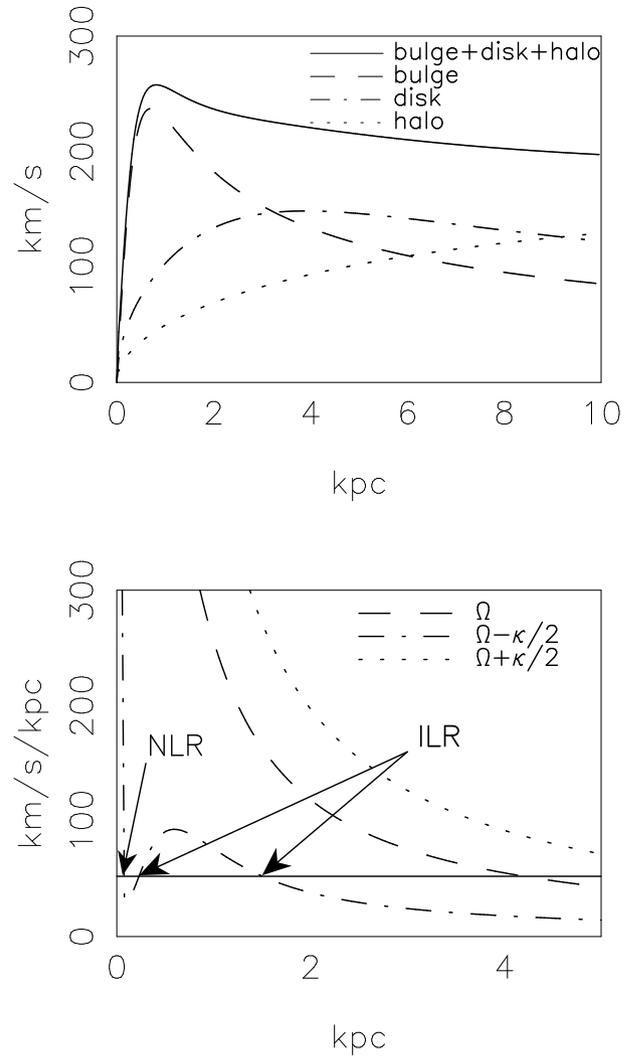}
 \end{center}
 \caption{Dependence of a distance from the galactic center on rotation velocities (top) and frequencies (bottom).
 In the upper graph, the solid, dashed, dash-dotted, and dot lines show rotation curves estimated by all components, the bulge, the disk, and the halo, respentively.
 In the lower graph, the horizontal solid line shows frequency of the stellar bar.
 The dashed, dash-dotted, and dotted lines show $\Omega$, $\Omega-\kappa /2$, and $\Omega + \kappa /2$, respectively.
 The ILR and the NLR points are shown by arrows.}\label{rotation_curve}
\end{figure}

Figure~\ref{rotation_curve} shows the rotation curve and the frequency obtained by our axisymmetric potential.
The corotation resonance is located at $4.2~{\rm kpc}$, and the ILRs are located at $0.2~{\rm kpc}$ and $1.5~{\rm kpc}$.
In addition to the ILRs, the NLR appears at $0.04~{\rm kpc}$ \citep{1998MNRAS.295..463F,2000ApJ...529..109F}.

Initial surface density profile of gas is exponential.
The scale length and the scale height of the gas disk is $7.0~{\rm kpc}$ and $400~{\rm pc}$, respectively, similarly to \citet{2008PASJ...60..667S}.
The gas disk is truncated at $10.5~{\rm kpc}$ which is slightly outer radius than the corotation resonance.
The mass of the initial gas disk is $5.4\times 10^{9}~M_{\odot}$.
The initial temperature of gas are set to be $10^4~{\rm K}$.
The initial metallicity is set to be a tenth of the solar metallicity in order to make initial formations of gas clouds mild and understand formations of gas clouds after the bar formation.

\subsection{Simulation methods}

In order to research the formation of star clusters, we performed high resolution simulations of gas dynamics in the nuclear region of a barred galaxy.
For the purpose, the numerical simulations are performed by the parallel Tree+GRAPE N-body/SPH code \textsc{ASURA} \citep{2008PASJ...60..667S}
with FAST time step scheme \citep{2010PASJ...62..301S}, the chemical evolution library (CELib) \citep{2017AJ....153...85S}, and density-independent SPH scheme \citep{2013ApJ...768...44S}.
For the hydrodynamics,
we take into account the radiative cooling with wide range of temperature ($10~{\rm K}<T<10^8~{\rm K}$), star formations from cold and dense gas, and heating by supernova (SN) feedback.
The star particle is spawned from the SPH particle
when the SPH particle satisfies the following conditions: (1) $T<100~{\rm K}$, (2) $n_{\rm H} > 100~{\rm cm^{-3}}$, and (3) $\nabla \cdot \bm{v} < 0$.
The mass of the newly formed stars is one third of the SPH mass.
The star particle is treated as a single stellar population (SSP) (see \citet{2008PASJ...60..667S}).
Here, the Kroupa's mass function is adopted for the star particle \citep{2001MNRAS.322..231K}.
Massive stars of which mass exceeds $8~M_{\odot}$ explode as Type II SNe and release energy of $10^{51}~{\rm erg}$.
Then, the SSP particles distribute thermal energy from the Type II SNe to the surrounding SPH particles.
Such simulations allow us to understand evolution of ISM, star formations from cold and dense gas, and formations of star clusters \citep{2009PASJ...61..481S,2012ApJ...746...26M,2019PASJ...71...19M}.

In our simulations, the initial number of SPH particles is $1,000,000$.
The mass of a SPH particle is $5.4\times 10^3~M_{\odot}$.
We adopt $5~{\rm pc}$ for the gravitational softening length.
We perform ${\rm RUN_{NSG}}$ which does not take self gravity of gas into account in order to study effects of self gravity of gas.
The resolution in our simulations is summarized in Tab.\ref{tab_parameters}.

\begin{table*}
  \tbl{Parameters of our simulations}{%
  \begin{tabular}{ccccc}
      \hline
	model	&	$N_{\rm SPH}$\footnotemark[$*$]	&	$m_{\rm SPH}$\footnotemark[$\dag$]	&	$\epsilon$\footnotemark[$\ddag$]	&	self gravity	\\
      \hline
       RUN				&	$1,000,000$				&	5.4 $\times 10^3~M_{\odot}$				&	$5~{\rm pc}$	&	$\surd$		\\
       ${\rm RUN_{NSG}}$	&	$1,000,000$	&	5.4 $\times 10^3~M_{\odot}$				&	$5~{\rm pc}$	&	$\times$		\\
      \hline
    \end{tabular}}\label{tab_parameters}
\begin{tabnote}
\footnotemark[$*$] The number of SPH particles.  \\ 
\footnotemark[$\dag$] The initial mass of SPH particles. \\
\footnotemark[$\ddag$] The gravitational softening length. \\ 
\end{tabnote}
\end{table*}

\section{Results}\label{results}

\subsection{Formations of Gas Clouds in the Central $x$-2 Ring}

\begin{figure}
 \begin{center}
  \includegraphics[width=80mm]{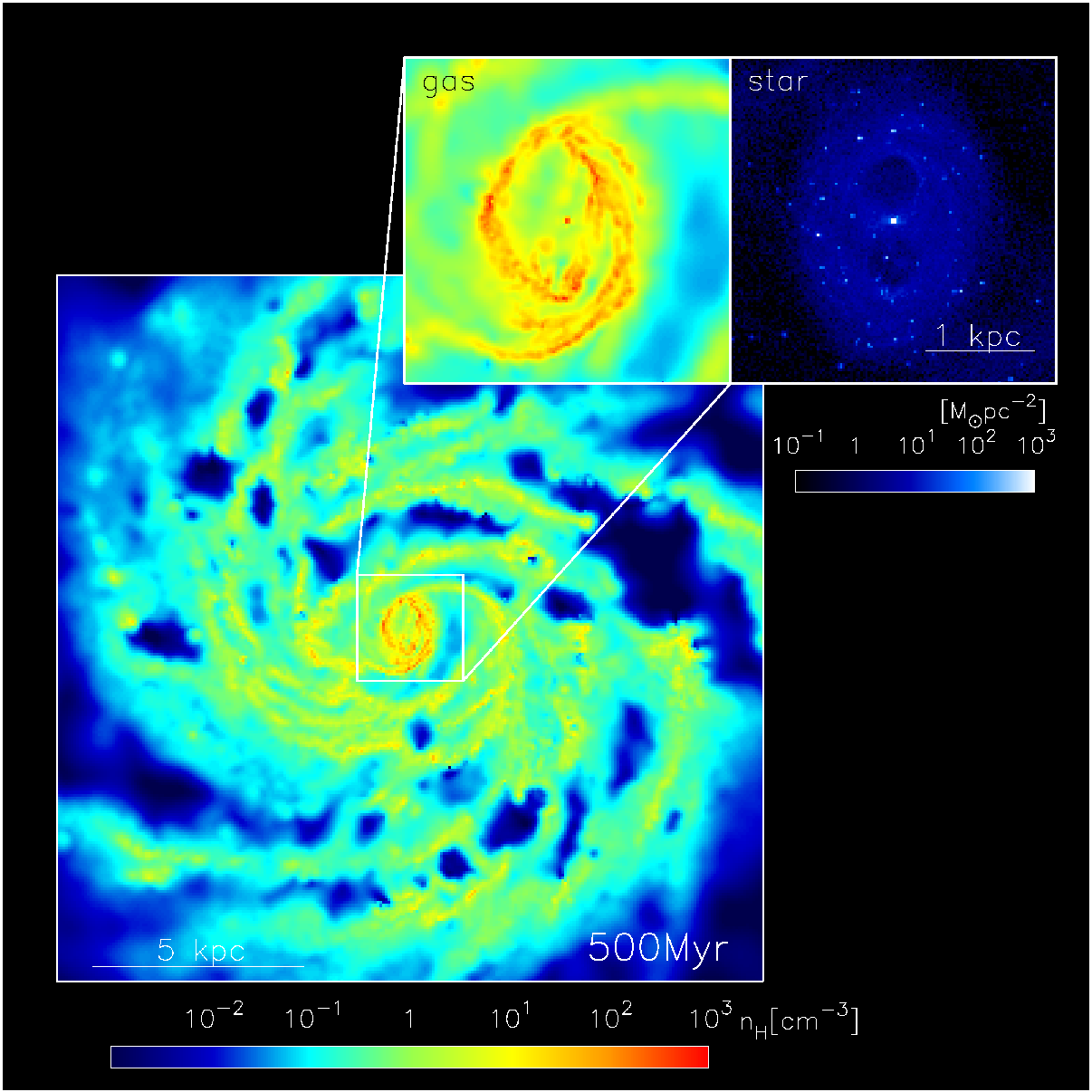}
 \end{center}
 \caption{A density map of the overall galaxy at $500~{\rm Myr}$.
 The upper small panels show surface density (left) and newly formed stars (right) within the galactic central $3~{\rm kpc}$.
 The stellar bar is along the horizontal axis of the figure.
 The centers of the upper right two panels are the galactic center.}\label{density_map}
\end{figure}

Figure~\ref{density_map} shows a gas density map in the simulated galaxy at $500~{\rm Myr}$.
As a bar grows from $200~{\rm Myr}$ to $500~{\rm Myr}$, spiral arms from the vicinity of the corotaion distance and gaseous ridges appear similarly to previous numerical simulations (e.g., \citet{1992MNRAS.259..345A}).
In the galactic central region, gas is accumulated less than the central $1~{\rm kpc}$ region because of gas inflow along the $x$-1 orbit.
Since the ILR barrier prohibits gas from falling promptly to the galactic center beyond ILR, a perpendicular oval ring, which is called the $x$-2 ring, is produced with respect to the bar.
In the $x$-2 ring, star formations take place, and star formation rate in the ring is $\sim 0.1-0.2~M_{\odot}~{\rm yr}^{-1}$.

The $x$-2 ring becomes dense after its formation.
In the ring, typical surface density and temperature of gas at $500~{\rm Myr}$ are $41~M_{\odot} ~ {\rm pc}^{-2}$ and $136~{\rm K}$, respectively.
Therefore, the Toomre Q value in the ring is

\begin{equation}
  Q= \frac{\kappa v_s}{\pi G \Sigma} = 0.87~\Big( \frac{v_s}{1.2~{\rm km/s}}\Big) \Big( \frac{41~M_{\odot}/{\rm pc}^2}{\Sigma} \Big) ,
\end{equation}

\noindent where $\kappa$ is the epicyclic frequency, $\Sigma$ is averaged gas surface density in the ring, and $v_s$ is sound speed.
The lower Q value than $1$ indicates that the $x$-2 gas ring is self gravitationally unstable.

\begin{figure}
 \begin{center}
  \includegraphics[width=80mm]{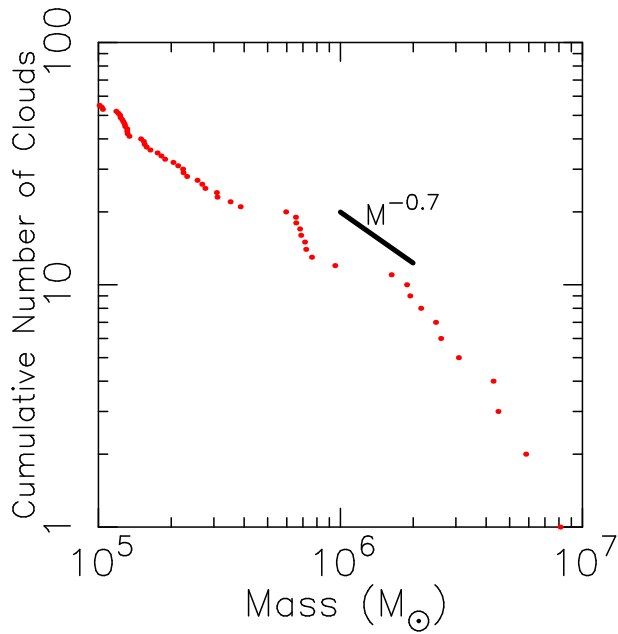}
 \end{center}
 \caption{Cumulative mass function of gas clouds identified by FOF algorithm in the $x$-2 orbit at $500~{\rm Myr}$.
 The horizontal and vertical axes show the cloud mass and the cumulative number of the gas clouds, respectively.}\label{CloudMassFunction}
\end{figure}

In RUN, the self gravitational instability facilitates fragmentations of the gas ring as shown in Fig.~\ref{density_map}.
Consequently, formations of gas clouds take place in the ring.
When the bar growth is completed at $500~{\rm Myr}$, $\sim 50$ gas clouds are identified in the $x$-2 ring.
Here, we use the Friends-Of-Friends (FOF) algorithm for the cloud identification \citep{Davis1985-zr}.
In the algorithm, $10~{\rm pc}$ is adopted for the linking length.
When the total mass exceeds $10^5~M_{\odot}$, we identify a group of SPH particles as a cloud.
Cumulative number of the identified clouds, which is given by

\begin{equation}
 N(>M)=\int _{M}^{M_{\rm max}} \phi (M)dM,
\end{equation}

\noindent in the $x$-2 ring at $500~{\rm Myr}$ is shown in Fig.~\ref{CloudMassFunction}.
The figure shows that the clouds have a power low mass function with $\phi (M) = dN/dM \propto M^{-1.7}$ which is consistent with a mass function of giant molecular clouds (GMCs) observed in neaby galaxies \citep{2020ApJ...893..135M}.
The maximum mass of the cloud is $8.1\times 10^6~M_{\odot}$.
Mean gas densities of the clouds range from $34~{\rm cm^{-3}}$ to $167~{\rm cm^{-3}}$.

Gas surface density in the $x$-2 ring declines with time due to consumption of gas by star formations or falls of gas clouds into the galactic center (see \S \ref{fueling}).
The decrease of the gas surface density increases the Q value in the ring.
As a result, the Q value exceeds $1$ after $\sim 600~{\rm Myr}$, which indicates that the gas ring becomes gradually stabilized due to the gas consumption.

\subsection{Impulsive Gas Fueling to the Galactic Center}~\label{fueling}

The formed gas clouds initially rotate in the $x$-2 ring orbit.
However, a cloud-cloud collision and/or energy injection into the gas ring by Type II SNe deviates some of gas clouds from the ring orbit.

\begin{figure}
 \begin{center}
  \includegraphics[width=80mm]{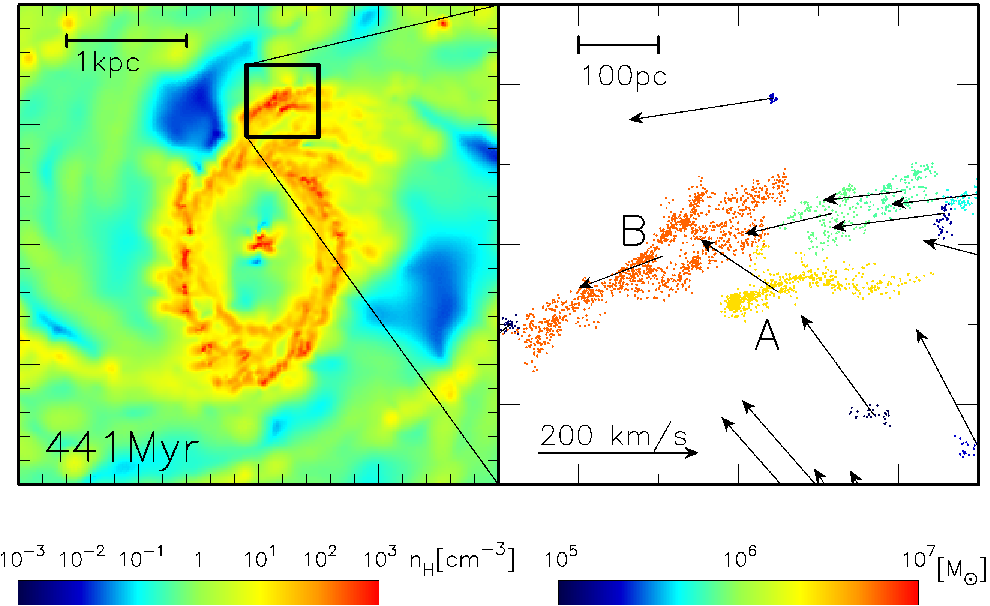}
 \end{center}
 \caption{Gas density map (left) and velocity map of identified gas clouds (right) at $441~{\rm Myr}$.
 The left panel shows the central $4~{\rm kpc}$ region.
 The right panel zooms in on the square region drawn in the left panel.
 In the right panel, arrows and colors show velocities and masses of the gas clouds.}\label{velocity}
\end{figure}

The left and right panels of Fig.~\ref{velocity} show gas density map at $441~{\rm Myr}$ in RUN and velocity map of clouds in the square drawn in the left panel, respectively.
In the right panel, three clouds are identified.
Hereafter, we refer to yellow and orange clouds as Cloud A and Cloud B, respectively.
The masses of Cloud A and Cloud B are $\sim 2\times 10^6~M_{\odot}$ and $\sim 3\times 10^6~M_{\odot}$, respectively.
The map shows that Could B moves in front of Cloud A.
Their relative velocity is $105~{\rm km/s}$ and these clouds collide soon.
The collision reduces the angular momentum of Cloud A and deviates Cloud A from the $x$-2 orbit.
Such cloud-cloud collisions take place near the apocenter distance, and consequently deviate some of gas clouds from the $x$-2 ring orbit.

\begin{figure}
 \begin{center}
  \includegraphics[width=80mm]{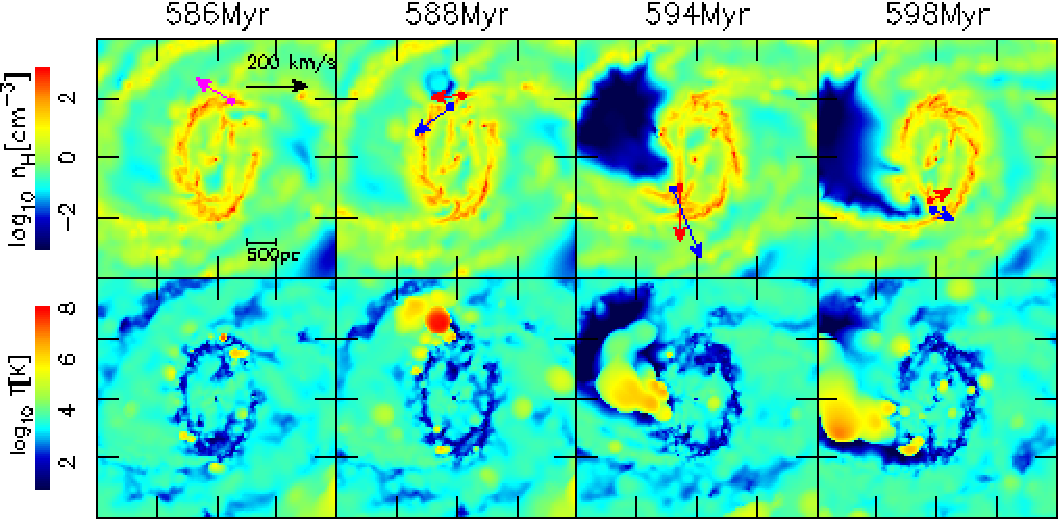}
 \end{center}
 \caption{Snapshots of gas density map (upper panels) and gas temperature map (lower panels) from $586~{\rm Myr}$ to $598~{\rm Myr}$.
 The size of each panel is $4~{\rm kpc}\times 4~{\rm kpc}$.
 The one cloud represented by the magenta dot in the upper panel at $586~{\rm Myr}$ is separated into two clouds after $588~{\rm Myr}$.
 The two separated clouds are represented by the red and blue dots in the upper panels after $588~{\rm Myr}$.
 Arrows attached to the dots show velocities of the clouds.
 }\label{TypeII}
\end{figure}

Similar deviation is also triggered by energy injection into the gas ring from Type II SNe.
Figure~\ref{TypeII} shows snapshots of gas density map and gas temperature map in the $x$-2 ring from $586~{\rm Myr}$ to $598~{\rm Myr}$.
We focus on evolution of one cloud which is represented by the magenta dot in the upper panel at $586~{\rm Myr}$.
Type II SNe generate near the cloud at $586~{\rm Myr}$.
The energy injection to gas by the Type II SNe heats the surrounding gas, thereby increasing the gas temperature more than $\sim 10^7~{\rm K}$.
The energy injection produces a cavity, which separates the cloud into two clouds.
The two clouds are represented by the red and blue dots in the upper panels after $588~{\rm Myr}$.
The motion of the two clouds are disturbed by Type II SNe.
As a result, the two clouds collide at $594~{\rm Myr}$. 
The collision deviates the one cloud from $x$-2 orbit.

\begin{figure}
 \begin{center}
  \includegraphics[width=80mm]{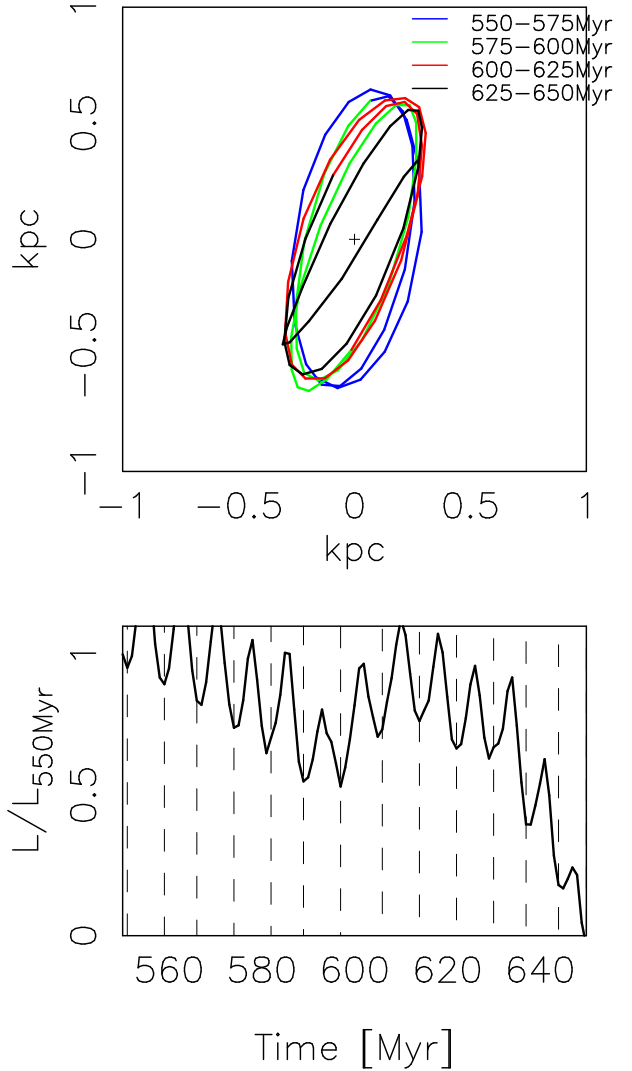}
 \end{center}
 \caption{The trajectory of one of deviated clouds (top) and time evolution of its angular momentum normalized by $550~{\rm Myr}$ (bottom).
In the upper panel, a dot represents the galactic center.
Blue, green, red, and black lines correspond to the trajectories from $550~{\rm Myr}$ to $575~{\rm Myr}$, from $575~{\rm Myr}$ to $600~{\rm Myr}$, from $600~{\rm Myr}$ to $625~{\rm Myr}$, and from $625~{\rm Myr}$ to $650~{\rm Myr}$, respectively.
In the lower panel, vertical dashed lines represent the epochs that the cloud is at the apocenter distance.}\label{trajectory}
\end{figure}

After the deviation, the trajectory of one deviated gas cloud is shown in the upper panel of Fig.~\ref{trajectory}.
We can see that the eccentricity of its orbit becomes higher during several rotations.
The cloud finally passes through the vicinity of the galactic center and falls into the galactic center.

The higher eccentric orbit is realized by decrease of the angular momentum.
The lower panel of Fig.~\ref{trajectory} shows the time evolution of normalized angular momentum of the cloud.
The angular momentum decreases from $610~{\rm Myr}$ to $650~{\rm Myr}$ and finally becomes nealy zero.
In this figure, we can also see the short-period oscillations of the angular momentum.
These oscillations result from the stellar bar potential.

\begin{figure}
 \begin{center}
  \includegraphics[width=80mm]{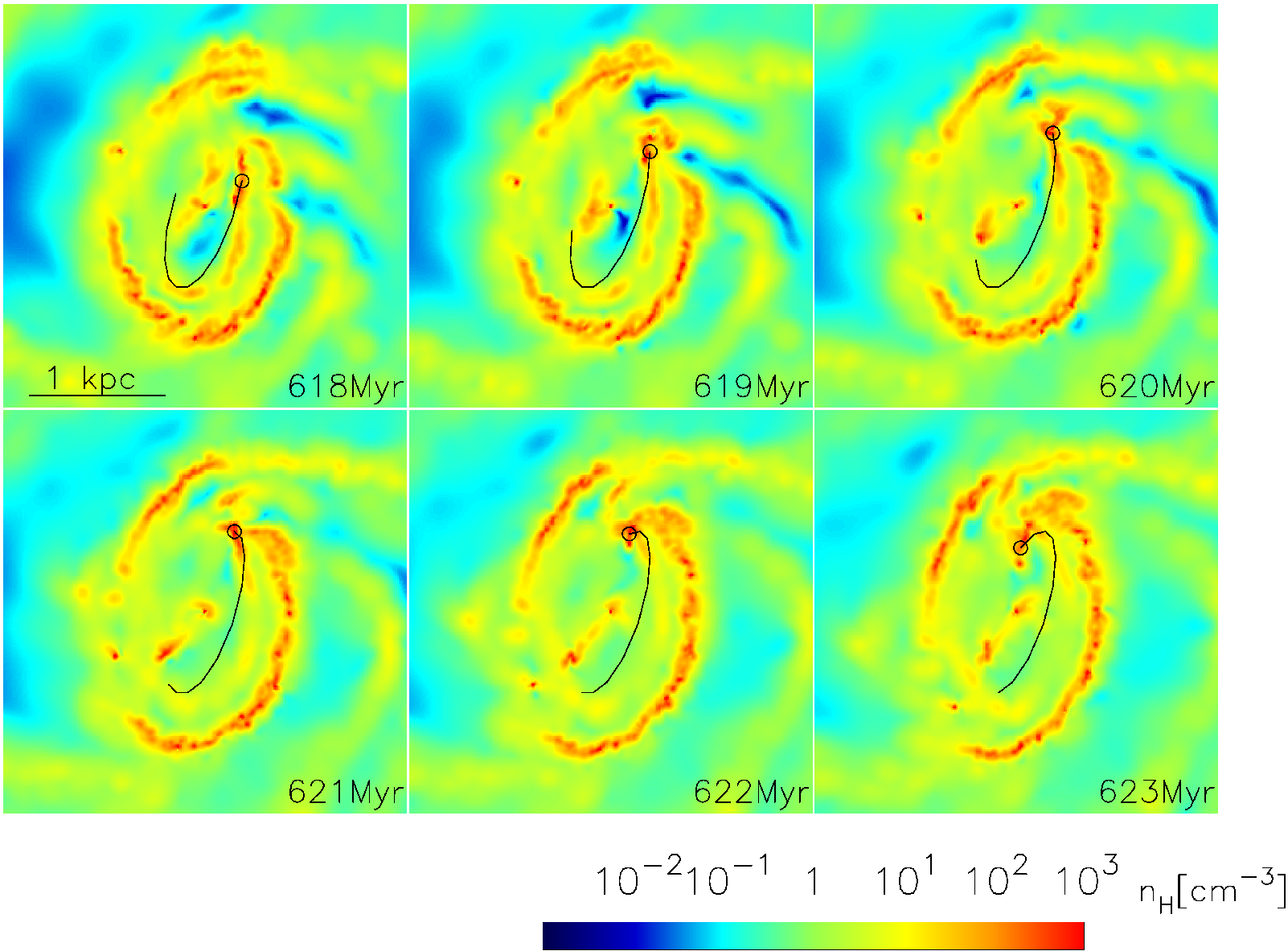}
 \end{center}
 \caption{Time evolution of position of one of deviated clouds from $618~{\rm Myr}$ to $623~{\rm Myr}$.
 The small circle shows the cloud position, and the solid line is its trajectory.}\label{cloud}
\end{figure}

Such decrease of the angular momentum of a deviated cloud is attributed to cloud-cloud collisions.
Snapshots in Fig.~\ref{cloud} show time evolution of gas surface density map from $618~{\rm Myr}$ to $623~{\rm Myr}$.
In the snapshots, the position of the deviated gas cloud is shown by a circle, and its trajectory is drawn by the solid line.
The figure shows that the deviated gas cloud collides with the other clouds, which still rotate in the $x$-2 orbit, near the apocenter distance.
Such collisions effectively reduce the angular momentum of the deviated cloud.
Since the angular momentum is reduced near the apocenter of the orbit, the eccentricity becomes higher without change of the apocenter distance.

\begin{figure}
 \begin{center}
  \includegraphics[width=80mm]{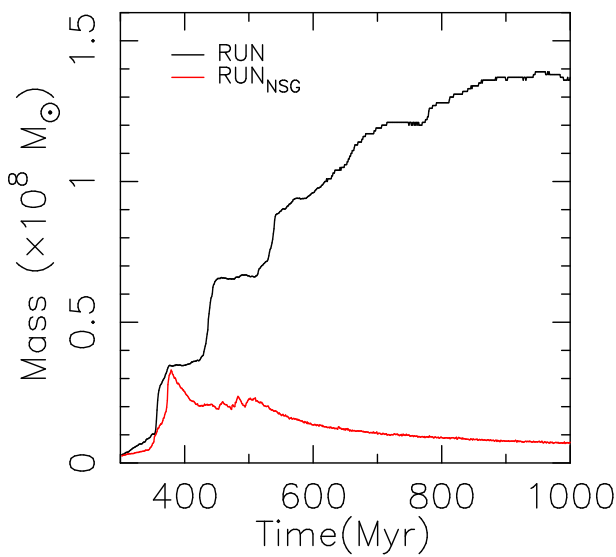}
 \end{center}
 \caption{Time evolution of the mass within the galactic central $10~{\rm pc}$.
 The black and red lines show the results of RUN and RUN$_{\rm NSG}$, respectively.}\label{centralmass}
\end{figure}

Falls of deviated clouds into the galactic center trigger the impulsive gas fueling to the galactic center.
Figure~\ref{centralmass} shows time evolution of the galactic central mass in the central $10~{\rm pc}$.
The increase of the central mass before $400~{\rm Myr}$ is caused by NLR as described in \S \ref{NLR}.
After $400~{\rm Myr}$, the central mass increases due to the impulsive gas fueling.
Falls of gas clouds take place at about $430~{\rm Myr}$, $530~{\rm Myr}$, $600~{\rm Myr}$, $630~{\rm Myr}$, $650~{\rm Myr}$, $700~{\rm Myr}$, $770~{\rm Myr}$, $810-840~{\rm Myr}$, and $940~{\rm Myr}$, thereby increasing the central mass by a few $10^7~M_{\odot}$ per fall of a gas cloud.
Finally, the central mass increases by $\sim 10^8~M_{\odot}$ after $400~{\rm Myr}$.

\section{Discussion}\label{discussion}

\subsection{Comparison with Observations}

In the central region of the nearby barred spiral galaxy M83, recent observations of dense molecular gas by the Atacama Large Millimeter/submillimeter Array (ALMA are able to identify individual molecular clouds \citep{2019ApJ...884..100H,2021MNRAS.505.4310C}.
In these observations, the molecular clouds seem to be distributed along the $x$-1 and $x$-2 orbits.
In addition, a few clouds are also observed in the inside of the $x$-2 orbit.

Our results suggest that formations of the molecular clouds observed along the $x$-2 orbit would take place because of self gravitational instability of the ring.
We also suggest that a few clouds observed inside the $x$-2 orbit would rotate originally in the $x$-2 orbit.
The cloud-cloud collisions in the ring and/or a large amount of energy injections by Type II SNe induce the deviation of gas clouds from the $x$-2 ring.
These deviated clouds lose their angular momentum due to the collisions near the apocenter distance with the other clouds which still rotate in the $x$-2 orbit.
Repeat of such collisions would make their orbit high eccentric, and finally such clouds would fall into the galactic center within $100~{\rm Myr}$.
These process would contribute to the AGN activity of M83 \citep{2016ApJ...824..107Y}.
Detailed observations of velocities of individual clouds would prove our scenario.
 
In this paper, we have shown formation of many massive dense clouds and several collisions between them in the nuclear dense gas ring by our numerical simulations.
Observational evidence of collision of massive dense clouds are reported in the central region in our Galaxy \citep{1994ApJ...429L..77H,2015PASJ...67..109T}.
In our Galaxy, many observational candidates of massive star formation indued by collision of molecular clouds are reported \citep{2021PASJ...73S...1F}.
Many simulations of cloud-cloud collisions are performed and show possibility of massive star formations by cloud-cloud collision \citep{2014ApJ...792...63T,2018PASJ...70S..58T,2018PASJ...70S..54S,2021PASJ...73S.385S,2023MNRAS.522.4972S}.
In these studies, the collision velocities is assumed to be in the range of $10-20~{\rm km/s}$.
Collision velocities found in our paper are much larger than these velocities.
It is very interesting to study possibility of massive star formation by collision simulations of massive clouds with such high collision velocity.

\subsection{Spatial Resolution}\label{resolution}

In the $x$-2 gas ring, the typical most unstable wavelength is estimated by

\begin{equation}
  \lambda = \frac{2\pi}{k} = \frac{2v_s^2}{G\Sigma} = 17~{\rm pc} \Big(\frac{v_s}{1.2~{\rm km/s}}\Big)^2 \Big( \frac{41~M_{\odot} {\rm pc}^{-2}}{\Sigma} \Big),
\end{equation}

\noindent from the Toomre instability analysis, where $v_s$ and $\Sigma$ are sound speed and averaged gas surface density in the ring, respectively.
In our simulations, $5~{\rm pc}$ is adopted for the gravitational softening length.
Since our spatial resolution is less than the most unstable wavelength in the ring, our simulations are eligible to follow fragmentations due to self gravitational instability in the $x$-2 gas ring.

\subsection{Gas fueling during the stellar bar growth phase}\label{NLR}

In Fig.~\ref{centralmass}, we should note that the galactic central mass increases by $\sim 5-7\times 10^7~M_{\odot}$ from $300~{\rm Myr}$ to $500~{\rm Myr}$ in RUN and RUN$_{\rm NSG}$.
Such mass increase is caused by gas inflow during the stellar bar growth phase.
During the stellar bar growth phase, a S-shaped gaseous structure forms within the $x$-2 ring in response to the bar growth.
Gas inflow to the galactic central region occurs along the S-shaped structure.
Similar gas structure and gas inflow can be seen in previous simulations in a barred galaxy \citep{1998MNRAS.295..463F,2000ApJ...529..109F}.
\citet{1998MNRAS.295..463F} performed numerical simulations with and without a galactic central SMBH.
In their simulation with a SMBH, a S-shaped gaseous structure appears and gas inflow occurs along the structure.
On the other hand, in their simulation without a SMBH, such structure does not appear.
They conclude that the gas inflow along the S-shaped structure results from the NLR.
Moreover, recent numerical simulations of barred galaxies have revealed that gas inflow effectively occurs during bar growth phase \citep{2015MNRAS.454.3641F,2020MNRAS.492.4500B}.
These results suggest that gas inflow along the S-shaped structure during the stellar bar growth phase would be attributed to the NLR and the stellar bar growth.

\subsection{Effects of self gravity}\label{self_gravity}

In ${\rm RUN_{NSG}}$, self gravity is not taken into account.
The red line in Fig.~\ref{centralmass} shows time evolution of mass in the galactic central $10~{\rm pc}$ in RUN$_{\rm NSG}$.
During the bar growth phase, S-shaped structures emerge due to the NLR, thereby triggering gas fueling to the galactic center.
Consequently, the galactic central mass evolves till $400~{\rm Myr}$ similarly to RUN.
The bar potential produces the $x$-2 ring.
Despite of the formation of the $x$-2 ring, gas fragmentations and formations of gas clouds do not occur in the ring unlike RUN.
Since cloud-cloud collisions do not take place, falls of gas clouds do not occur in ${\rm RUN_{NSG}}$.
The mass within the galactic $10~{\rm pc}$, therefore, does not increase after the bar grows.
These results suggest that the NLR triggers gas fueling to the galactic center during the bar growth phase and self gravity in the $x$-2 ring plays important role in transporting gas from $x$-2 ring to the galactic center after the bar grows.
In other words, fragmentations of gas and cloud-cloud collisions in the $x$-2 ring are essential to induce impulsive gas fueling.

\subsection{Effects of metallicity}\label{metallicity}

In our simulations, the initial metallicity is set to be 0.1 of the solar metallicity.
The abundance of metallicity does not affect formation of the $x$-2 ring.
On the other hand, it affects instability of the $x$-2 ring after its formation since the time scale of the radiative cooling depends on the metallicity.
If metallicity is the solar metallicity, the $x$-2 ring cools more rapidly and tends to be unstable in more earlier stage.
Therefore, our scenario should take place in not only barred galaxy observed at high redshift (e.g., \citet{2023ApJ...945L..10G}) but also nearby barred galaxies.

\section{Conclusion}\label{conclusion}

Our numerical simulations of interstellar medium in barred galaxies have demonstrated that self-gravitational instability in the $x$-2 ring induces formations of gas clouds.
Although formed gas clouds initially rotate in the $x$-2 orbit, cloud-cloud collisions and/or energy injections into the ring by Type II SNe make some of the gas clouds deviate from the $x$-2 orbit.

The deviated clouds collide with the other clouds, which rotate in the $x$-2 orbit, near the apocenter distance.
Such collision effectively reduces the angular momentum and makes the orbital eccentricity high rather than reducing the apocenter distance.
Several collisions finally trigger falls of gas clouds, and consequently gas fueling to the galactic center takes place.
Our results insist that these processes would play an important role in evolution of the galactic nuclear region, for examples AGN activities, growth of SMBHs, and nuclear starburst in barred galaxies.

\bigskip
The authors thank the anonymous referee for providing us with helpful comments.
HM is funded by JSPS KAKENHI Grant (21K03621).
AH is funded by JSPS KAKENHI Grant (19K03923).
TRS is funded by JSPS KAKENHI Grant (21K03614,21K03633,22H01259,22K03688).
Numerical computations were carried out on Cray XC50 at Center for Computational Astrophysics (CfCA), National Astronomical Observatory of Japan,
and numerical analysis were also carried out on computers at CfCA.
The authors are grateful to Dr. Yoshimasa Watanabe, Prof. Yoshiaki Taniguchi, and Prof. Kenji Bekki for helpful discussions. 
The authors are also grateful to Mr. Yuya Kudo, Mr. Fuga Asari, and Mr. Takumi Inoue affiliated with National Institute of Technology, Asahikawa College, for supporting our research.

\appendix
\section*{Incorporation of the Coriolis force}

Our simulations adopt a rotating coordinate system as described in \S \ref{models}.
In a rotating coordinate system, incorporation of the Coriolis force into a simulation code is essential.
In order to calculate the Coriolis force, the velocity of a particle has to be used.
How to adopt the velocity is one of the difficult problems in the leap frog integration scheme. 

We calculate acceleration at $n$-th step and velocity at $(n+1/2)$-th step as follows:

\begin{equation}
 \left( \begin{array}{c}
  a_x ^{n} \\
  a_y ^{n} \\
 \end{array} \right)
 =
  2\Omega _{\rm bar}
 \left( \begin{array}{c}
  v_y ^{n-1/2} \\
  -v_x ^{n-1/2} \\
 \end{array} \right)
  + \Omega  _{\rm bar}^2
 \left( \begin{array}{c}
  x ^{n} \\
  y ^{n}  \\
 \end{array} \right)
 + 
 \left( \begin{array}{c}
  f_{x}^{n} \\
  f_{y}^{n} \\
 \end{array} \right)
\end{equation}

\begin{equation}
 \left( \begin{array}{c}
  v_x ^{n+1/2} \\
  v_y ^{n+1/2} \\
 \end{array} \right)
 =
 \left( \begin{array}{c}
  v_x ^{n-1/2} \\
  v_y ^{n-1/2} \\
 \end{array} \right)
 +
 \left( \begin{array}{c}
  a_x ^{n} \\
  a_y ^{n} \\
 \end{array} \right)
 \Delta t
\end{equation}

\noindent Here, the first, the second, and the third terms in the right side of the equation show the Coriolis force, the centrifugal force, and the gravitational force, respectively.

\begin{figure}
 \begin{center}
  \includegraphics[width=80mm]{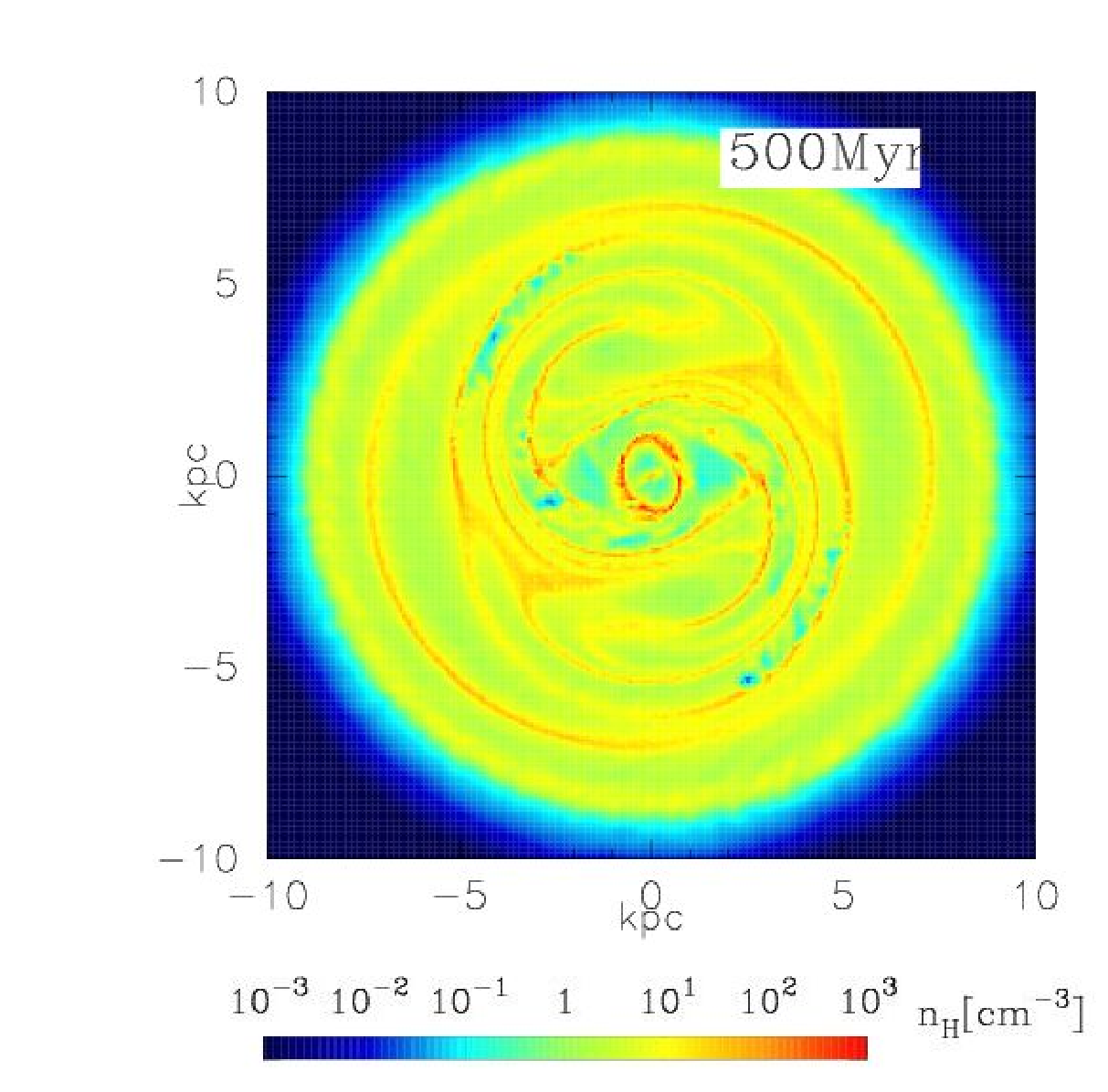}
 \end{center}
 \caption{Gas density map at $500~{\rm Myr}$ in RUN$_{\rm NSG}$.}\label{appendix}
\end{figure}

Figure~\ref{appendix} shows the density map at $500~{\rm Myr}$ in RUN$_{\rm NSG}$.
This figure indicates that the $x$-2 ring, gaseous ridge along the bar, and spiral arms from the bar end can be reproduced clearly.
These structures are good agreement with those in previous simulations (e.g., \citet{1992MNRAS.259..345A,1992MNRAS.258...82W}).
Since the Coriolis force plays an important role in reproducing these structures, these results suggest that our treatment of the Coriolis force is appropriate to follow gas dynamics in our galactic scale simulations.

\end{document}